\def\ARXIV{1}
\RequirePackage{fix-cm}
\ifdefined\ARXIV
  \documentclass[final,5p,times,twocolumn]{elsarticle}
  \newcommand{\wideleftmargin}{0pt}
  \newcommand{\widerightmargin}{0pt}
  \newenvironment{widefigure}{\begin{figure*}[!t]}{\end{figure*}}
  \newenvironment{widetable}{\begin{table*}[!t]}{\end{table*}}
\else
  \documentclass[preprint,12pt,a4paper]{elsarticle}
  \newcommand{\wideleftmargin}
    {-\dimexpr(\paperwidth-\textwidth)/2-13.3mm\relax}
  \newcommand{\widerightmargin}
    {-\dimexpr(\paperwidth-\textwidth)/2-13.3mm\relax}
  \newenvironment{widefigure}{\begin{figure}[!t]}{\end{figure}}
  \newenvironment{widetable}{\begin{table}[!t]}{\end{table}}
\fi

\usepackage{amssymb}
\usepackage{amsmath}
\usepackage{algorithm}
\usepackage{algpseudocode}
\usepackage{afterpage}
\usepackage{hyperref}
\hypersetup{hidelinks}
\usepackage{caption}
\usepackage{booktabs}
\usepackage{changepage}
\usepackage[scaled=0.95]{inconsolata}
\usepackage{siunitx}

\DeclareMathOperator{\Cov}{Cov}
\DeclareMathOperator{\Var}{Var}
\newcommand{\eps}{\varepsilon}
\newcommand{\N}{\mathcal{N}}
\journal{SoftwareX}

\begin{document}
\renewcommand{\labelenumii}{\arabic{enumi}.\arabic{enumii}}

\begin{frontmatter}

\title{Burn-in-Free Simulation of VARMA Time Series}

\author[hi]{Kristján Jónasson\corref{cor1}}
\cortext[cor1]{Corresponding author}
\ead{jonasson@hi.is}
\address[hi]{University of Iceland, Reykjavík, Iceland}

\begin{abstract}
  Varmapack is a software package for efficient, exact simulation of VARMA time
  series without a burn-in period. For stationary models, Varmapack can generate
  initial states and innovations from their joint stationary distribution.
  Alternatively, the user can supply initial states, in which case innovations
  are generated from their conditional distribution. The core C library has
  interfaces to R, Python, and Matlab. It also offers VARMAX simulation and
  computation of autocovariances and correlations, spectral radii, and impulse
  responses. Varmapack automatically selects between vector-Yule-Walker and
  state-space methods for covariance computation and uses level-3 BLAS
  operations for efficient generation of multiple replicates. Benchmarks on
  several platforms show substantial performance gains over existing simulation
  software, ranging from several-fold to more than three orders of magnitude for
  the packages and models considered. Varmapack is open source and publicly
  available through GitHub.
\end{abstract}


\end{frontmatter}
\ifdefined\ARXIV
\thispagestyle{plain}
\fi

\ifdefined\ARXIV
\else
\section*{Metadata}
\label{sec:metadata}

\begin{widetable}
\begin{tabular}{|l|p{6.5cm}|p{6.5cm}|}
\hline
\textbf{Nr.} & \textbf{Code metadata description} & \textbf{Metadata} \\
\hline
C1 & Current code version & \\
\hline
C2 & Permanent link to code/repository used for this code version & \\
\hline
C3 & Permanent link to Reproducible Capsule & \\
\hline
C4 & Legal Code License & \\
\hline
C5 & Code versioning system used & \\
\hline
C6 & Software code languages, tools, and services used & \\
\hline
C7 & Compilation requirements, operating environments \& dependencies & \\
\hline
C8 & If available link to developer documentation/manual & \\
\hline
C9 & Support email for questions & \\
\hline
\end{tabular}
\caption{Code metadata (mandatory)}
\label{tab:metadata}
\end{widetable}
\fi

\section{Motivation and significance}
\label{sec:introduction}
Vector autoregressive moving-average (VARMA) time series models date back to the
last part of the 20th century, as a generalization of scalar ARMA, and vector AR
models and MA models (VAR and VMA); see \cite{duker2025review} for a review of
the development and early papers. A straightforward way to simulate VARMA time
series for a specified model is to start with random (or zero) data, and use the
model to compute future values, injecting random innovations as needed. A
problem with this procedure is that an initial, possibly substantial, part of
the simulated series must be discarded for the results to approximate the
correct distribution. This \emph{burn-in} can be avoided by making use of the
exact combined distribution of innovations and the values of the series to start
the simulation.

Already in 1978 an algorithm for exact ARMA simulation was described
\cite{mcleod1978simulation}, in fact very similar to the scalar case of the
current algorithm. In 1987 exact VARMA simulation using a state-space
formulation was described \cite{barone1987simulation}, and the following year a
revised formulation was described \cite{shea1988simulation}, essentially quite
similar to the current one, although neither was accompanied by software or
simulation-study results.

Two R packages offer exact scalar ARMA simulation: ltsa, which uses
Durbin--Levinson recursions and requires a preliminary computation of the
autocovariance sequence \cite{mcleod2007ltsa,golub1996matrix}, and
ts.extend, which takes the ARMA coefficients directly and can
condition on specified observations \cite{oneill2021tsextend}. The latter can
be expensive for long series because it constructs their full covariance
matrix. Two other R packages, MTS and beyondWhittle,
simulate VARMA models by direct forward recursion after a burn-in segment has
been discarded \cite{tsay2022mts,meier2026beyondwhittle}. In Python,
Statsmodels provides two relevant classes: VAR, for direct
forward simulation of VAR models, and the state-space VARMAX, which
can perform exact stationary simulation but requires conversion of coefficient
matrices to its internal parameter vector \cite{seabold2010statsmodels}.

In 2008 the present author published two papers on VARMA likelihood with missing
values \cite{jonasson2008method,jonasson2008algorithm}, which included, in
addition to theoretical results, a suite of Matlab programs to analyse and
simulate VARMA models: TOMS Algorithm 878. The simulations described in these
papers are burn-in-free (spin-up-free). The current work builds on the 2008
articles. Their notation is largely followed and the forward recursion is
identical, but the generation of the startup values differs slightly: Appendix C
of \cite{jonasson2008method} draws the time series states first and the
innovations conditionally on them, whereas in the current algorithm this order
is reversed. The current package also adds VARMAX simulation and computation of
spectral radii, covariances, and impulse responses as detailed in the next
sections. The Matlab programs have been translated to C, and interfaces to
Python, R, and Matlab are provided.

\section{Mathematical background}
\label{sec:mathematical-background}

This section derives the formulae used by the software for VARMA simulation and the computation of derived quantities. The corresponding formulae for VARMAX simulation are provided in the repository documentation \cite{varmax-startup}.

\subsection{Covariance structure and exact initialization}
\label{sec:covariance-initialization}
Let $x_t \in \mathbb{R}^r$ follow a zero-mean, stationary $\mathrm{VARMA}(p,q)$
process:
\begin{equation}
  \label{eq:varma}
  x_t
  = \sum_{j=1}^{p} A_j x_{t-j}
  + \eps_t
  + \sum_{j=1}^{q} B_j \eps_{t-j},
\end{equation}
where the $\eps_t$ are $\mathcal{N}(0, \Sigma)$ innovations (white noise
shocks), and the $x_t$ are the time series states. Let $\Gamma_j$ denote the
lagged state covariances, $\Cov(x_t, x_{t-j})$, and $C_j$ the lagged mixed
covariances, $\Cov(x_t, \eps_{t-j})$. The $C_j$ are given by $C_0 = \Sigma$ and
\begin{equation}
  C_j = \sum_{k=1}^{j} A_k C_{j-k} + B_j\Sigma \quad (j=1,2,\ldots).
\end{equation}
Here $A_i$ and $B_j$ are taken as $0$ for $i>p$ and $j>q$, respectively. Note
that there is a mistake in eq. (7) in \cite{jonasson2008method} where the last
term should be $B_j\Sigma$, not $B_q\Sigma$. $\Gamma_0,\ldots,\Gamma_{p-1}$ are
obtained by solving the vector-Yule-Walker equations as described in Appendix B
of \cite{jonasson2006efficient} (where they are called $S_0,\ldots,S_{p-1}$).
The remaining $\Gamma_j$ are obtained recursively with $B_0=I$ as
\begin{equation}
  \label{eq:gamma-recursion}
  \Gamma_j = \sum_{k=1}^{p} A_k\Gamma_{j-k}
    + \sum_{k=j}^{q} B_k C_{k-j}^T.
  \quad j=p,p+1,\ldots,
\end{equation}

Let $h \geq \max(p,q)$ and define the stacked vectors
$$
  x_{1:h} = (x_1^T, \ldots, x_h^T)^T
  \quad\text{and}\quad
  \eps_{1:h} = (\eps_1^T, \ldots, \eps_h^T)^T.
$$
Let $S$, $C$, and $\overline{\Sigma}$ denote, respectively, $\Var x_{1:h}$,
$\Cov(x_{1:h}, \eps_{1:h})$, and $\Var \eps_{1:h}$. The two stacked vectors are
jointly Gaussian with distribution
\begin{equation}
  \label{eq:joint-startup}
  \begin{bmatrix}
    x_{1:h} \\
    \eps_{1:h}
  \end{bmatrix}
  \sim
  \mathcal{N}\left(
    0,
    \begin{bmatrix}
      S & C \\
      C^T & \overline{\Sigma}
    \end{bmatrix}
  \right).
\end{equation}
It follows that
\begin{equation}
  \label{eq:x-given-eps}
  x_{1:h}\mid\eps_{1:h} \sim \mathcal{N}(m,Q),
  \quad m = C\overline{\Sigma}^{-1}\eps_{1:h},
  \quad Q = S - C\overline{\Sigma}^{-1}C^T,
\end{equation}
and
\begin{equation}
  \label{eq:eps-given-x}
  \eps_{1:h}\mid x_{1:h} \sim \mathcal{N}(e,R),
  \quad e = C^T S^{-1}x_{1:h},
  \quad R = \overline{\Sigma} - C^T S^{-1}C.
\end{equation}

The covariance matrix $S$ is block Toeplitz with $(i,j)$-block
\begin{equation}
  S_{ij} =
  \begin{cases}
    \Gamma_{i-j} & \text{for } i\geq j\\
    \Gamma_{j-i}^T & \text{for } i<j,\\
  \end{cases}
\end{equation}
$C$ is block lower-triangular with $(i,j)$-block
\begin{equation}
  C_{ij} =
  \begin{cases}
    C_{i-j} & \text{for } i\geq j\\
    0 & \text{for } i<j\\
  \end{cases}
\end{equation}
and $\overline{\Sigma}$ is block diagonal with all $h$ blocks equal to $\Sigma$. All three
matrices are $rh \times rh$.
Simpler formulae for $m$ and $Q$ are obtained by setting $\Psi =
C\overline{\Sigma}^{-1}$, which, like $C$, is a lower-triangular block matrix.
Block $(i,j)$ is $\Psi_{i-j}$, which can be computed by setting $\Psi_0 = I$ and
computing $\Psi_1,\ldots,\Psi_{h-1}$ with the recurrence:
\begin{equation}
  \label{eq:psi-j}
  \Psi_j = \sum_{k=1}^{j} A_k\Psi_{j-k} + B_j
\end{equation}
where as before $A_k$ and $B_j$ are taken as $0$ for $k>p$, $j>q$. Then
\begin{equation}
  \label{eq:x-given-eps-psi}
  m = \Psi\eps_{1:h}
  \quad\text{and}\quad
  Q = S - \Psi\overline{\Sigma} \Psi^T.
\end{equation}
The $\Psi_j$ are the impulse-response coefficients discussed in Section
\ref{sec:derived-quantities}.

As an alternative to the vector-Yule-Walker equations, define the state vector
\begin{equation*}
  u_t = (x_t^T,\ldots,x_{t-p}^T,\eps_t^T,\ldots,\eps_{t-q}^T)^T.
\end{equation*}
It has the state-space representation $u_t=Fu_{t-1}+J\eps_t$, where the first
block row of $F$ is
\begin{equation*}
  [A_1\ \cdots\ A_p\ 0\mid B_1\ \cdots\ B_q\ 0],
\end{equation*}
the remaining block rows shift the states and innovations, and $J$ has identity
blocks in the positions for $x_t$ and $\eps_t$. The stationary covariance
matrix $P=\Var(u_t)$ then solves the discrete-time Lyapunov equation
\begin{equation}
  \label{eq:lyapunov}
  P = FPF^T + J\Sigma J^T.
\end{equation}
The required $\Gamma_j$ and $C_j$ are in the first block row of $P$
\cite{barone1987simulation,shea1988simulation}.

\subsection{Derived model quantities}
\label{sec:derived-quantities}
\paragraph{Autocovariances}
Given model parameters, lagged theoretical covariances $\Gamma_k$ up to any lag
may be computed as described in Section \ref{sec:covariance-initialization}, and
given observed states $x_0,\ldots,x_{n-1}$, lagged data covariances may be
computed using
\begin{equation}
  \label{eq:gammak}
  \widehat{\Gamma}_k = \frac{1}{n}\sum_{t=k}^{n-1}
  (x_t-\bar{x})(x_{t-k}-\bar{x})^T.
\end{equation}
Equation \eqref{eq:gammak} uses maximum-likelihood normalization and is biased.
To partially correct for the bias, $n$ can be replaced with $n-k$ in the
denominator (if the true mean were used instead of the sample mean, the
correction would give an exactly unbiased formula).

\emph{Autocorrelations} are easily obtained from autocovariances by dividing by
the product of the corresponding lag-zero marginal standard deviations.

\paragraph{Spectral radii}
The autoregressive spectral radius $\rho$ is the maximum absolute eigenvalue of
the autoregressive companion matrix
$$
\mathcal{A} =
\begin{bmatrix}
  A_1    & A_2    & A_3    & \cdots & A_p \\
  I      & 0      & 0      & \cdots & 0   \\
  0      & I      & 0      & \ddots & \vdots \\
  \vdots & \ddots & \ddots & \ddots & 0   \\
  0      & \cdots & 0      & I      & 0
\end{bmatrix}.
$$
A model is stationary when $\rho<1$. The moving-average spectral radius
$\rho_{\textrm{MA}}$ is the spectral radius of the moving-average companion
matrix (whose first block row is $-B_1,\ldots,-B_q$). It is an invertibility
diagnostic: when it is less than one, the moving-average polynomial $B(L)$ has a
convergent inverse,
$$
  B(L)^{-1} = I+\Delta_1L+\Delta_2L^2+\ldots.
$$
The model can then be written as an equivalent infinite pure VAR model
$$
  x_t=\Pi_1x_{t-1}+\Pi_2x_{t-2}+\ldots+\varepsilon_t,
$$
where
$$
  I-\Pi_1L-\Pi_2L^2-\ldots
  = (I+\Delta_1L+\Delta_2L^2+\ldots)A(L).
$$

\paragraph{Impulse responses}
The impulse response matrix $\Psi_j$ maps a change in the innovation at time $t$
to the resulting change in the process at time $t+j$. For a stationary model,
$$
  x_t=\sum_{j=0}^{\infty}\Psi_j\varepsilon_{t-j}.
$$
Thus the $\Psi_j$ are the coefficients of the infinite VMA representation.
With $\Psi_0=I$, they satisfy

$$
  \Psi_j=B_j+\sum_{i=1}^{\min(p,j)}A_i\Psi_{j-i}, \quad j\geq1,
$$
where $B_j=0$ for $j>q$. Orthogonalized impulse responses can also be computed as
$$
  \Theta_j=\Psi_jL,
$$
where $LL^T=\Sigma$ and $L$ is the lower Cholesky factor of a positive-definite
innovation covariance matrix \cite{lutkepohl2005multiple,keating1996structural}.

\section{Software description}
\label{sec:software-description}
\subsection{Language interfaces}
\label{sec:language-interfaces}

The core of the present software package is a C library for exact simulation of
VARMA and VARMAX time series and computation of the derived quantities discussed
in Section \ref{sec:derived-quantities}. The public header file
\texttt{varmapack.h} serves as a compact reference for the C API: all
user-facing functions are declared there, with comments describing the role of
each parameter. Simulation uses Randompack \cite{jonasson2026randompack}. The
library is thread safe, and its diagnostic error state is local to each thread.

The package also contains interfaces to Python, R, and Matlab. The Python and R
interfaces closely match one another: Python defines a \texttt{Model} class and
R defines an R6 \texttt{VarmapackModel}, constructed with
\texttt{varmapack\_model}. In both, model parameters are stored in the object,
and simulation and most derived quantities are provided as corresponding
methods. The Matlab interface is instead functional, with model parameters
passed directly to functions in the \texttt{varmapack} namespace.

\subsection{VARMA simulation}
\label{sec:varma-simulation}

There are two possibilities to start VARMA simulation with Varmapack: (a) by
drawing both shocks $\varepsilon_t$ and states $x_t$ from the exact joint
distribution of $(x,\varepsilon)$ for the initial segment $t=0,\ldots,h-1$,
where $h=\max(p,q)$, and (b) by specifying $h \geq\max(p,q)$ initial values of
the series and drawing the first $h$ shocks from the conditional distribution of
$(\varepsilon|x)$. After starting, both cases simply run the sequence forward by
drawing innovations from $\mathcal{N}(0,\Sigma)$ and computing states using the
model. Using the vector-Yule-Walker covariance calculation, the following
algorithm fills $rn\times M$ matrices $E$ and $X$ with $M$ replicates of shocks
and corresponding states for case (a).
\begin{algorithm}[H]
\caption*{VARMA simulation with random start}
\begin{algorithmic}[1]
\item Find the vector-Yule-Walker right-hand sides $G_j$ as in
\cite[Appendix B]{jonasson2006efficient}.
\item Solve the vector-Yule-Walker equations for $S_{ij}$.
\item Compute $\Psi$ using \eqref{eq:psi-j}.
\item Compute $Q$ with \eqref{eq:x-given-eps-psi}.
\item Fill $E$ with independent $\varepsilon_t\!\sim\!\mathcal N(0,\Sigma)$.
\item Fill the first $rh$ rows of $X$ with $\mathcal N(0,Q)$ draws.
\label{it1:X}
\item Set $X_{1:rh,:}:=X_{1:rh,:}+\Psi E_{1:rh,:}$,
c.f. \eqref{eq:x-given-eps} and \eqref{eq:x-given-eps-psi}.
\item Use \eqref{eq:varma} to compute remaining states.\label{it:remaining}
\end{algorithmic}
\end{algorithm}
\vspace{-\baselineskip}
Case (b) is similar, except that instead of steps 3–7 the starting innovations
are drawn using \eqref{eq:eps-given-x}. For this to work, $S$ must be positive
definite; a sufficient (but not necessary) condition for that is that $\Sigma$
be positive definite.

For a stationary model, both (a) and (b) are possible, and the simulated series
will have the correct distribution from the first generated term, so that no
burn-in segment needs to be discarded. For a nonstationary model, starting
values must be supplied. For a pure VAR model ($q=0$), the recurrence simply
runs forward from those values. If MA terms are present, startup innovations are
drawn from their theoretical distribution, conditional on constraints imposed by
the model and supplied states. $\Sigma$ must be positive semidefinite and is
allowed to be singular except for nonstationary models with MA terms.

VARMA simulation may have a fixed or time-dependent mean path $\mu_t$. The
recursion is then applied to the centered series $x_t-\mu_t$. Explicit mean
paths are not currently supported for VARMAX simulation.

For case (a), Algorithm 878 \cite{jonasson2008algorithm} takes the reverse,
mathematically equivalent approach: it draws $x_{1:h}$ from $\N(0,S)$ and then
draws $\eps_{1:h}$ from the conditional distribution in
\eqref{eq:eps-given-x}. This is more complex and expensive than the current
procedure.

The covariance setup may use either the vector-Yule-Walker equations or the
state-space formulation \eqref{eq:lyapunov}, solved with the SLICOT routine
\texttt{SB03MD} \cite{benner1999slicot}. The former is faster for smaller $r$
and the latter for larger $r$, with the crossover depending on $p$ and $q$.
Varmapack selects between them automatically using empirically determined
cutoffs.

It may happen that the conditional covariance $Q$ is singular even for
well-behaved models. For example, for the two-dimensional VAR(1) model
$x_t=Ax_{t-1}+\eps_t$ with $a_{ij}=0.1$ and $\Sigma=2I$, $Q$ is singular with
$q_{ij}=1/24$ for all $i,j$. Varmapack handles such cases through Randompack’s
multivariate normal routine, which uses pivoted Cholesky factorization
(\texttt{dpstrf}) when ordinary Cholesky factorization fails.

\subsection{VARMAX simulation}

With VARMAX simulation, startup states must be provided: $x_t$ for
$t=0,\ldots,h-1$, where $h\geq\max(p,q,s-1)$, as well as the whole sequence of
exogenous values $z_t$, $t=0,\ldots,n-1$. As for nonstationary VARMA, $\Sigma$
must be positive definite, and the startup innovations are then drawn from their
theoretical distribution, conditional on constraints imposed by the model and
the supplied states and exogenous sequence. The derivation is provided in the
repository documentation \cite{varmax-startup}. The interfaces name the exogenous
coefficient argument \texttt{C}; mathematically it contains the blocks
$D_1,\ldots,D_s$.
When more than $\max(p,q,s-1)$
values of $x_t$ and corresponding $z_t$ are available, they should be supplied
to improve the information used to draw the innovations.

\subsection{Portability and verification}
\label{sec:verification}

The Varmapack C library has been successfully installed and tested on macOS,
Linux (x86-64 and ARM64), and Windows, using the C compilers gcc, clang, icx,
nvc, and MSVC, and the Fortran compilers gfortran, ifx, and nvfortran, with BLAS
from Accelerate, OpenBLAS, and MKL.

Test suites are provided for the C library and its Python, R, and Matlab
interfaces. Together, the tests exercise every supported operation with the
applicable parameter configurations, including multiple replicates, supplied
starting values, fixed and time-dependent means, and exogenous series. They also
check output shapes, reproducibility with seeded random-number generators, edge
cases, and error reporting.

The C test suite additionally checks numerical results against an independent
Matlab reference implementation over all named test cases, comparing
autocovariances, spectral radii, impulse responses, and simulated paths using
the same Randompack streams where applicable. Most reference functions derive
from the thoroughly tested Matlab code in Algorithm~878, while newer functions
were developed independently of the C implementation. The suite also
cross-checks the vector-Yule-Walker and Lyapunov covariance solvers and tests
generated models, positive-semidefinite covariance paths, nonfinite and
overflow-prone inputs, and the low-level error state.

\begin{widefigure}
  \centering
  \makebox[\textwidth][c]{\includegraphics[width=365pt]{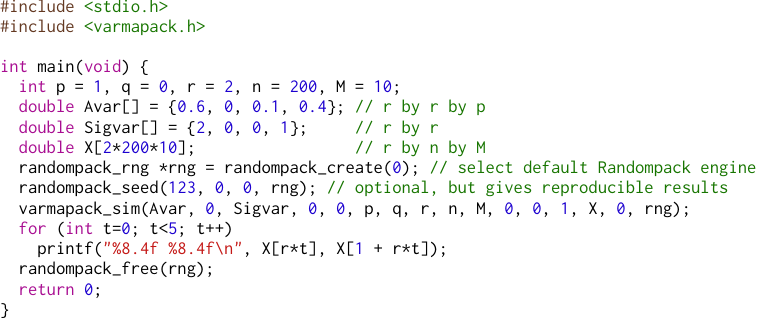}}
  \caption{C example. This program generates 10 replicates of length 200 from
    a bivariate VAR(1) model and prints the first five values of the first
    replicate. Error checking is omitted.}
  \label{fig:c-example}
\end{widefigure}

\begin{widefigure}
  \begin{adjustwidth}
    {\wideleftmargin}
    {\widerightmargin}
  \centering
  \includegraphics[width=\linewidth]{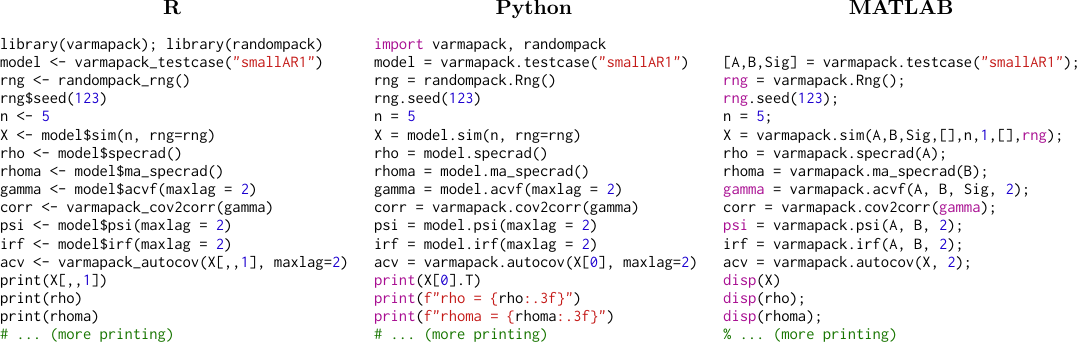}
  \end{adjustwidth}
  \caption{The same VARMA calculation in the three high-level interfaces.}
  \label{fig:language-example}
\end{widefigure}

\begin{figure}[!b]
  \centering
  \includegraphics[width=0.77\linewidth]
    {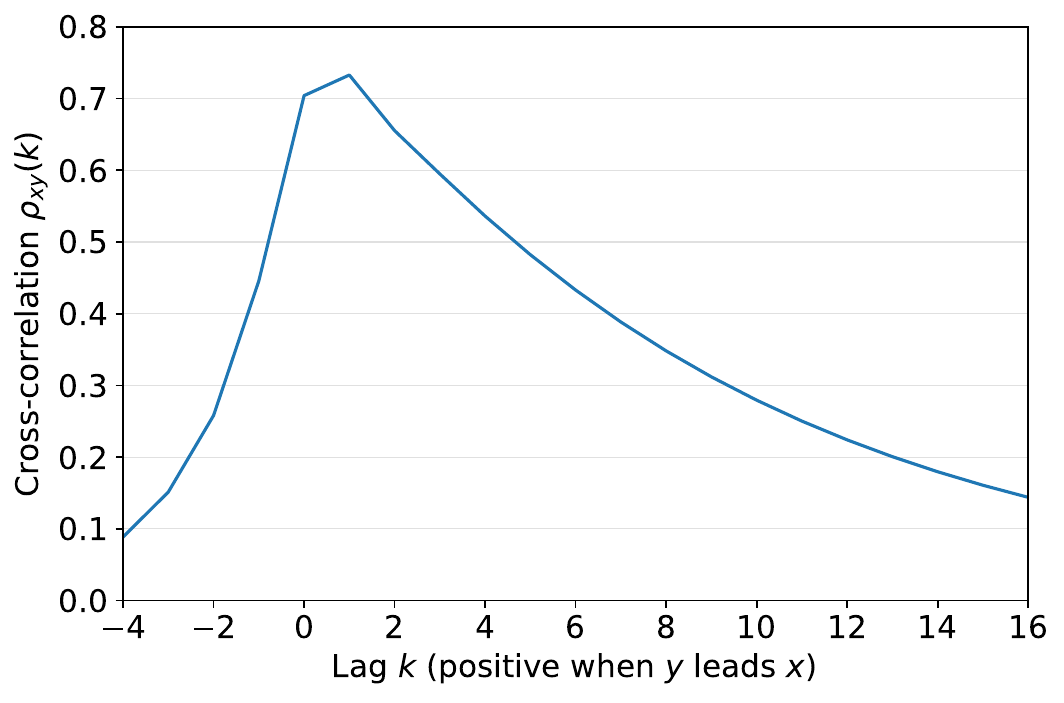}
  \caption{The theoretical cross-correlation
    $\rho_{xy}(k)=\operatorname{Corr}(x_t,y_{t-k})$. At positive lags,
    $y_t$ leads $x_t$.}
  \label{fig:bivariate-cross-correlation}
\end{figure}

\begin{figure}[!b]
  \centering
  \includegraphics[width=\linewidth]
    {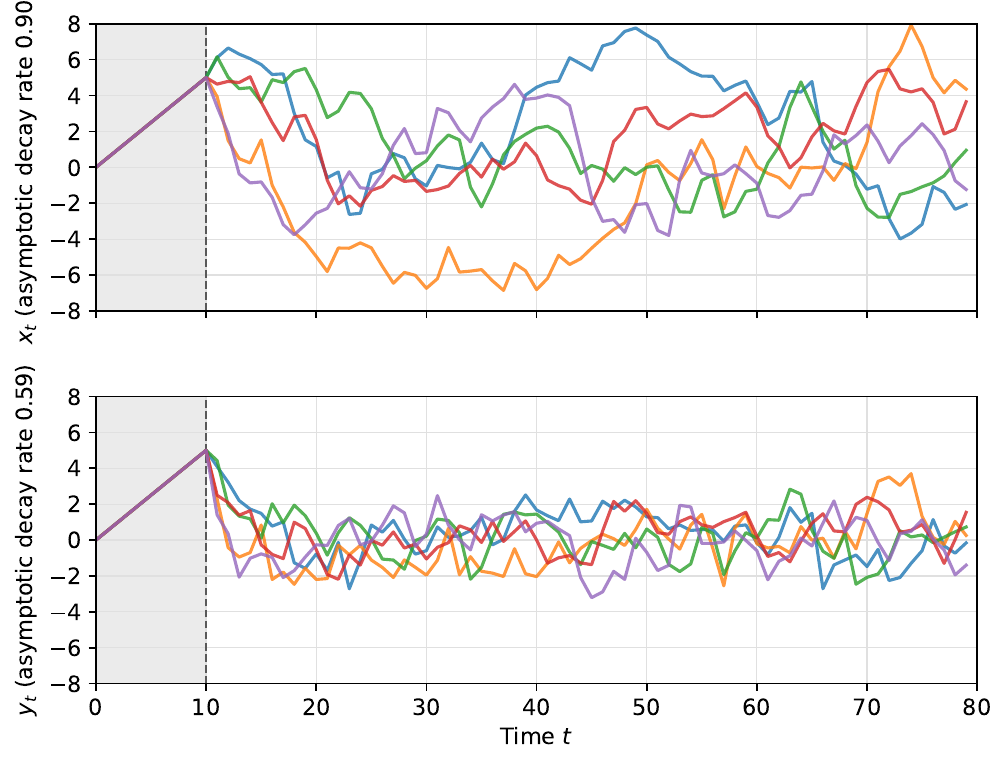}
    \caption{Five simulated replicates of a bivariate VARMA(2,1) model with
      fixed initialization path. The $x_t$ component is more persistent than
      $y_t$; on close inspection, the paths also reveal the positive
      cross-correlation between the components, with $y_t$ leading $x_t$.}
  \label{fig:bivariate-paths}
\end{figure}

\afterpage{%
  \begin{widetable}
    \centering
    \caption{Simulation timings for Varmapack and Statsmodels VARMAX on
      five platforms, in nanoseconds per simulated value.}
    \label{tab:simulation-timings}
    \begin{adjustwidth}
    {\wideleftmargin}
    {\widerightmargin}
  \centering
  \setlength{\tabcolsep}{2pt}
  \setlength{\cmidrulewidth}{\lightrulewidth}
  \fontsize{8.5}{10.5}\selectfont
  \begin{tabular*}{\linewidth}{@{\extracolsep{\fill}}
      l
      c@{\hspace{1em}}
      S[table-format=1.0]
      S[table-format=1.0]
      S[table-format=1.0]
      c@{\hspace{1em}}
      S[table-format=1.3]
      c@{\hspace{1em}}
      S[table-format=2.1]
      S[table-format=3.0]
      c@{\hspace{1em}}
      S[table-format=2.1]
      S[table-format=3.0]
      c@{\hspace{1em}}
      S[table-format=2.1]
      S[table-format=4.0]
      c@{\hspace{1em}}
      S[table-format=2.1]
      S[table-format=3.0]
      c@{\hspace{1em}}
      S[table-format=2.1]
      S[table-format=3.0]
    }
    \toprule
    & & & & & & & & \multicolumn{2}{c}{Mac-M4} && \multicolumn{2}{c}{Spark} && \multicolumn{2}{c}{Xeon} && \multicolumn{2}{c}{Core-i5} && \multicolumn{2}{c}{Windows} \\
    \cmidrule{9-10}
    \cmidrule{12-13}
    \cmidrule{15-16}
    \cmidrule{18-19}
    \cmidrule{21-22}
    Model && {$p$} & {$q$} & {$r$} && {$\rho$} && {Ours} & {VARMAX} && {Ours} & {VARMAX} && {Ours} & {VARMAX} && {Ours} & {VARMAX} && {Ours} & {VARMAX} \\
    \midrule
    \texttt{tinyARMA} && 1 & 1 & 1 && 0.400 && 6.2 & 241 && 9.9 & 252 && 23.4 & 745 && 13.1 & 368 && 9.2 & 302 \\
    \texttt{smallAR2} && 2 & 0 & 2 && 0.974 && 9.3 & 150 && 5.4 & 163 && 12.9 & 482 && 7.8 & 243 && 5.7 & 187 \\
    \texttt{mediumAR} && 1 & 0 & 3 && 0.516 && 7.0 & 103 && 4.4 & 110 && 9.8 & 334 && 6.6 & 168 && 4.2 & 130 \\
    \texttt{mediumARMA1} && 3 & 3 & 3 && 0.960 && 11.4 & 325 && 6.2 & 209 && 12.1 & 642 && 9.9 & 309 && 5.7 & 228 \\
    \texttt{largeAR} && 5 & 0 & 7 && 0.957 && 5.3 & 525 && 5.8 & 371 && 9.8 & 1072 && 8.5 & 562 && 5.0 & 397 \\
    \bottomrule
  \end{tabular*}
\end{adjustwidth}

  \end{widetable}

  \begin{widetable}
    \centering
    \caption{Simulation timings for Varmapack and other packages on Mac-M4.}
  \label{tab:package-timings}
    \begin{adjustwidth}
    {\wideleftmargin}
    {\widerightmargin}
  \centering
  \setlength{\tabcolsep}{2pt}
  \setlength{\cmidrulewidth}{\lightrulewidth}
  \fontsize{8.5}{10.5}\selectfont
  \begin{tabular*}{0.85\linewidth}{@{\extracolsep{\fill}}
      l
      c@{\hspace{1em}}
      S[table-format=1.0]
      S[table-format=1.0]
      S[table-format=1.0]
      c@{\hspace{1em}}
      S[table-format=1.3]
      c@{\hspace{1em}}
      S[table-format=2.1]
      c@{\hspace{1em}}
      S[table-format=2.1]
      S[table-format=5.0]
      S[table-format=2.0]
      c@{\hspace{1em}}
      S[table-format=3.0]
      S[table-format=3.0]
      c@{\hspace{1em}}
      S[table-format=4.0]
    }
    \toprule
    & & & & & & & & \multicolumn{1}{c}{C} && \multicolumn{3}{c}{R} && \multicolumn{2}{c}{Python} && \multicolumn{1}{c}{Matlab} \\
    \cmidrule{9-9}
    \cmidrule{11-13}
    \cmidrule{15-16}
    \cmidrule{18-18}
    Model && {$p$} & {$q$} & {$r$} && {$\rho$} && {Varmapack} && {Varmapack} & {MTS} & {ts.extend} && {VARMAX} & {VAR} && {varm} \\
    \midrule
    \texttt{tinyAR} && 1 & 0 & 1 && 0.500 && 6.0 && 9.7 & 9625 & 64 && 231 & 40 && 2276 \\
    \texttt{tinyARMA} && 1 & 1 & 1 && 0.400 && 6.3 && 10.6 & 15576 & 64 && 246 & {\textendash} && {\textendash} \\
    \texttt{smallAR1} && 1 & 0 & 2 && 0.200 && 4.4 && 6.5 & 5203 & {\textendash} && 144 & 68 && 1326 \\
    \texttt{smallARMA2} && 1 & 2 & 2 && 0.500 && 9.1 && 7.6 & 11974 & {\textendash} && 157 & {\textendash} && {\textendash} \\
    \texttt{mediumAR} && 1 & 0 & 3 && 0.516 && 7.0 && 5.8 & 3531 & {\textendash} && 104 & 61 && 907 \\
    \texttt{mediumARMA2} && 3 & 3 & 3 && 0.993 && 11.2 && 8.3 & 14743 & {\textendash} && 289 & {\textendash} && {\textendash} \\
    \texttt{largeAR} && 5 & 0 & 7 && 0.957 && 5.3 && 8.0 & 5855 & {\textendash} && 538 & 120 && 1577 \\
    \texttt{largeARMA} && 3 & 3 & 7 && 0.980 && 6.5 && 8.4 & 7107 & {\textendash} && 521 & {\textendash} && {\textendash} \\
    \bottomrule
  \end{tabular*}
\end{adjustwidth}

  \end{widetable}
}

\section{Illustrative examples}
\label{sec:illustrative-examples}
This section illustrates the C, R, Python, and Matlab interfaces and presents a
bivariate VARMA example demonstrating graphing of the theoretical
cross-correlation and multiple simulated paths.

\paragraph{Example 1: C interface}
The example in Fig.~\ref{fig:c-example} is adapted from an example in the
package's C readme file.

\paragraph{Example 2: R, Python, and Matlab interfaces}
Figure~\ref{fig:language-example} compares the three high-level interfaces. The
examples create the same test model, simulate a short series, and compute
spectral radii, autocovariances and correlations, and standard and
orthogonalized impulse responses. With the same Randompack seed, all three
produce the simulated values
$$
\begin{array}{rrrrr}
  1.3923 & -1.1619 & -1.7641 & 0.1571 & 0.6110\\
  2.0876 & -1.9181 & 1.2999 & -3.5218 & -2.7306
\end{array}
$$
and the spectral radii $\rho = 0.200$ and $\rho_{\textrm{MA}} = 0$.

\paragraph{Example 3: Bivariate VARMA model}
This zero-mean VARMA(2,1) example computes and plots multiple simulated paths
together with the theoretical lagged cross-correlation between the two
components:
$$
\begin{aligned}
  A_1 &= \begin{pmatrix}0.75&0.05\\0&0.50\end{pmatrix}, &
  A_2 &= \begin{pmatrix}0.13&0\\0&0.05\end{pmatrix}, \\
  B_1 &= \begin{pmatrix}0.40&0.15\\0.05&0.20\end{pmatrix}, &
  \Sigma &= \begin{pmatrix}1&0.99\\0.99&1\end{pmatrix}.
\end{aligned}
$$
All five replicates use the fixed initialization path $x_t=y_t=t/2$ for
$t=0,\ldots,10$.
The results are plotted in Figs.~\ref{fig:bivariate-cross-correlation}
and~\ref{fig:bivariate-paths}.

\section{Impact}
\label{sec:impact}
\subsection{Research applications}
Simulation of time series is widely used in the evaluation of estimation and
forecasting methods, bootstrap procedures, power analyses, and teaching. For
multivariate series, an important requirement is to preserve both temporal and
cross-variable dependence, whether in methodological studies or in applications
such as the generation of long synthetic environmental and ocean-wave records
from comparatively limited observations
\cite{cai2011multivariate,guanche2013climate,
  azimmohseni2015simulation,valsamidis2022simulation}.

Varmapack provides this capability for general VARMA models, rather than being
restricted to univariate series or pure VAR models. Unlike simulation methods
that rely on a burn-in period, it generates observations from the correct
distribution from the outset. Its C implementation is designed for efficient
generation of multiple replicates, using level-3 BLAS operations for the main
matrix computations. As shown in Section~\ref{sec:performance}, this also gives
substantial performance gains over existing general-purpose software.

The review article \cite{duker2025review} discusses how VARMA models can give
more parsimonious representations of the underlying process than VAR models and
may improve predictive accuracy; see also \cite{wilms2021bigtime} for empirical
evidence. Tiao and Box \cite{tiao1981modeling} also discuss the potential
advantages of VARMA over VAR and VMA.

As described in Section~\ref{sec:varma-simulation}, Varmapack supports two forms
of exact initialization. For stationary models, the starting states and
innovations can be drawn jointly from their stationary distribution, giving the
simulated series its exact unconditional stationary distribution from the first
generated term. Alternatively, starting states can be supplied by the user, with
the corresponding initial innovations drawn from their conditional distribution.
The resulting series then has the exact distribution conditional on the supplied
states. Neither approach requires choosing a burn-in length or discarding an
initial segment of the simulation.

\subsection{Performance}
\label{sec:performance}
Performance is assessed with the provided \texttt{TimeSimulate} benchmark
programs. In this section, Varmapack is compared with simulation functions in
several other packages for selected models on an M4 Mac, and it is also
benchmarked across five platforms alongside Statsmodels VARMAX.

The other packages have differing capabilities. ts.extend provides
exact simulation but only for univariate ARMA models; MTS supports
VARMA models but uses a burn-in of 200 observations by default; Statsmodels
VARMAX supports exact VARMA simulation through a state-space
representation; and VAR and Matlab's \texttt{varm} are restricted to
VAR models and require burn-in, for which 200 values were discarded in the
benchmarks. For low persistence this is sufficient, but highly persistent models
need longer burn-in. For the \texttt{mediumARMA2} model, for example, MTS with
its default burn-in gives a variance about 7\% too low for the first retained
value; if the same model without its MA terms is simulated with Statsmodels'
VAR or Matlab's \texttt{varm}, the discrepancy is about 6\%.

Table~\ref{tab:package-timings} compares Varmapack with the other packages on an
M4 Mac. The reported timings are median nanoseconds per simulated value over 11
runs of the timing program, using 1000 replicates of 100 time steps. Python and
Matlab Varmapack timings were essentially identical to C and are omitted from
the table. Varmapack is about 600–1800 times faster than MTS, 6–7 times
faster than ts.extend, 15–100 times faster than VARMAX, 6–23
times faster than VAR, and 130–380 times faster than \texttt{varm}.
Gains of this magnitude can substantially expand the scale and scope of feasible
simulation studies, allowing orders of magnitude more replicates and much
broader exploration of models and parameter settings.

Table~\ref{tab:simulation-timings} compares Varmapack with Statsmodels
VARMAX on five platforms. As in Table~\ref{tab:package-timings}, the
reported timings are median nanoseconds per simulated value over 11 runs of the
timing program, using 1000 replicates of 100 time steps. The Varmapack C and
Python timings were essentially identical and are combined in the columns headed
``Ours’’. Varmapack is 15–109 times faster than VARMAX across the
selected models and platforms. On the M4 Mac clang and Accelerate were used,
MSVC and MKL on Windows, and gcc and OpenBLAS on the other platforms.

\section{Conclusions and availability}
\label{sec:conclusion}
Varmapack provides efficient, burn-in-free simulation of VARMA models, with C,
R, Python, and Matlab interfaces and supporting tools for model analysis.
Planned extensions include likelihood evaluation and estimation, structured
VARMA specifications, and treatment of missing observations.

The Varmapack source code, examples, benchmark programs, and test suites are
available at \url{https://github.com/jonasson2/varmapack}. The Python and R
interfaces are also distributed through PyPI and CRAN, respectively; the Matlab
interface is included in the GitHub repository.

\bibliographystyle{elsarticle-num}
\bibliography{paper}

\end{document}